\documentclass{article}
\usepackage{spconfa4,amsmath,graphicx, xcolor, amssymb}
\usepackage{bm}
\usepackage{acronym}
\usepackage{hyperref}
\usepackage{booktabs}
\usepackage{tikz}
\usetikzlibrary{positioning, fit, arrows.meta, calc}
\usepackage{subcaption}
\usepackage{siunitx}
\usepackage{balance}
\usepackage{cite}

\usepackage{fancyhdr}
\acrodef{FAD}{Fréchet Audio Distance}
\acrodef{MMD}{Maximum Mean Discrepancy}
\acrodef{WER}{Word Error Rate}
\acrodef{CER}{Character Error Rate}
\acrodef{PCC}{Pearson Correlation Coefficient}
\acrodef{SRCC}{Spearman Rank Correlation Coefficient}

\newcommand{\featmodel}{f}
\newcommand{\idxLayer}{l}
\newcommand{\featDim}{d}

\newcommand{\speechFrame}{\mathbf{x}}
\newcommand{\embedding}{\mathbf{e}}

\newcommand{\setInp}{\mathcal{X}}
\newcommand{\setEmb}{\mathcal{E}}
\newcommand{\setRef}{\mathcal{E}_{\mathrm{r}}}
\newcommand{\setTest}{\mathcal{E}_{\mathrm{t}}}
\newcommand{\numEmbR}{M}
\newcommand{\numEmbT}{N}

\newcommand{\muRef}{\bm{\mu}_{\mathrm{r}}}
\newcommand{\muTest}{\bm{\mu}_{\mathrm{t}}}
\newcommand{\SigmaRef}{\bm{\Sigma}_{\mathrm{r}}}
\newcommand{\SigmaTest}{\bm{\Sigma}_{\mathrm{t}}}

\newcommand{\tr}{\operatorname{tr}}
\newcommand{\FAD}{\mathrm{FAD^2}}
\newcommand{\FADformula}{%
  \left\| \muRef - \muTest \right\|_2^2
  + \tr\!\left[
      \SigmaRef + \SigmaTest
      - 2\!\sqrt{\SigmaRef \SigmaTest}
    \right]
}
\newcommand{\MMD}{\mathrm{MMD}}
\newcommand{\kernel}{k}
\newcommand{\distR}{p_{\mathrm{r}}}
\newcommand{\distT}{p_{\mathrm{t}}}

\title{OBJECTIVE INTELLIGIBILITY PREDICTION USING DISTANCE METRICS ON SPEECH FOUNDATION MODEL REPRESENTATIONS}
\name{Lyonel Behringer, Andreas Brendel\thanks{This work was partially supported by the Free State of Bavaria in the DSGenAI project.}}
\address{Fraunhofer Institute for Integrated Circuits (IIS), 
Erlangen, Germany}
\begin{document}
\ninept
\maketitle

\thispagestyle{fancy}
\pagestyle{empty} 

\begin{abstract}
High-dimensional representations of pretrained speech foundation models have proven beneficial for objective speech quality and intelligibility prediction. While existing work on neural intelligibility prediction usually leverages such representations for task-specific fine-tuning, in this work we evaluate the usefulness of such representations for intelligibility prediction without any further training. We conduct a layer-wise analysis of multiple speech foundation models, correlating various embedding distances with subjective intelligibility scores. The results show that embeddings extracted from Whisper speech recognition models are best suited, with the last encoder and decoder layers yielding the best correlations when using the Fréchet Audio Distance. Notably, the evaluated distances outperform classical intelligibility metrics and are more robust than Word and Character Error Rates. Further, correlations improve with increasing size of the Whisper model from which embeddings are extracted.
\end{abstract}


\section{Introduction}
Successful voice communication requires received speech to be intelligible. Speech technologies such as hearing aids, speech codecs, or speech enhancement systems are dedicated to preserve or even improve intelligibility~\cite{2nd_clarity_enhancement_challenge, yi2026_lm_loss_for_ulb_speech_coding, oshaughnessay_speech_enhancement_review}.
To ensure sufficient intelligibility after processing by such technologies, adequate assessment methods are required. As a gold standard, speech intelligibility assessment is done with subjective tests~\cite{schmidt1995intelligibility}. However, this is costly and time intensive.

To evaluate intelligibility of speech technologies more efficiently, objective metrics designed to imitate human hearing may be used. 
While such metrics have traditionally been
based on classical signal processing~\cite{french1947factors, steeneken1980physical, ASA_ANSI_S3.5-1997,  elhilali2003spectro, TaalSTOI, jensen2016_estoi} or probabilistic and hybrid automatic speech recognition (ASR) systems~\cite{KARBASI_asr_based_SIP}, neural approaches based on highly generalizing speech foundation models (SFMs) have gained popularity recently. High-dimensional representations of pretrained SFMs such as wav2vec 2.0~\cite{NEURIPS2020_wav2vec2} or Whisper~\cite{radford2023robust} have seen successful use as basis for training intelligibility prediction models~\cite{cuervo2024_sfm_intelligibility, zezario_mosanetplus}. However, it is unclear whether such prediction models are robust to out-of-distribution input signals.
Further, the use of word error rate (WER) via neural ASR models has become popular, but works using such techniques are usually limited to the English language~\cite{behringer2026assessingimpactnoisespeech}. Moreover, the corresponding model outputs are not differentiable, limiting their use as an intelligibility-based loss function.

\usetikzlibrary{arrows.meta, fit, backgrounds}

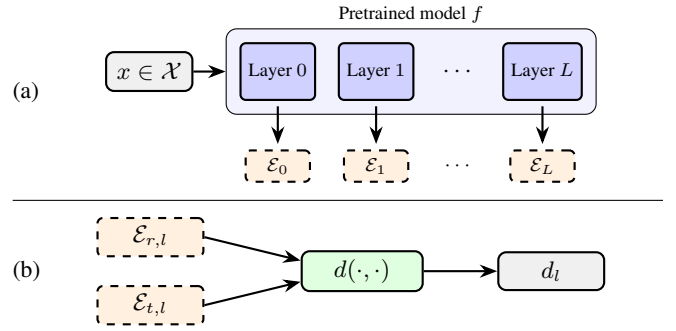
\begin{figure}[t]
\centering

\begin{minipage}[c]{0.04\linewidth}
(a)
\end{minipage}%
\begin{minipage}[c]{0.96\linewidth}
\centering
\begin{tikzpicture}[
    >=Stealth, thick,
    signal/.style={draw, rounded corners=3pt, fill=gray!12,
        inner sep=4pt, font=\small, align=center},
    layer/.style={draw, fill=blue!18, rounded corners=2pt,
        minimum width=0.85cm, minimum height=0.75cm, font=\scriptsize, align=center},
    embset/.style={draw, dashed, rounded corners=2pt, fill=orange!12,
        minimum width=0.85cm, minimum height=0.40cm, font=\scriptsize, align=center},
    modelbox/.style={draw, rounded corners=4pt, fill=blue!5, inner sep=5pt}
]

\node[layer] (l0) at (0,   0) {Layer $0$};
\node[layer] (l1) at (1.3, 0) {Layer $1$};
\node[font=\small] (ld) at (2.4, 0) {$\cdots$};
\node[layer] (lL) at (3.5, 0) {Layer $L$};

\begin{scope}[on background layer]
    \node[modelbox, fit=(l0)(l1)(ld)(lL),
          label={[font=\scriptsize,inner sep=2pt]above:Pretrained model $f$}] (m) {};
\end{scope}

\node[signal] (sig) at (-1.7, 0) {$x \in \mathcal{X}$};
\draw[->] (sig.east) -- (m.west);

\node[embset] (e0) at (0,   -1.25) {$\mathcal{E}_{0}$};
\node[embset] (e1) at (1.3, -1.25) {$\mathcal{E}_{1}$};
\node[font=\scriptsize]  at (2.4, -1.25) {$\cdots$};
\node[embset] (eL) at (3.5, -1.25) {$\mathcal{E}_{L}$};

\foreach \l in {0,1,L} {
    \draw[->, shorten >=2pt, shorten <=2pt] (l\l.south) -- (e\l.north);
}
\end{tikzpicture}
\end{minipage}

\vspace{0.6em}
\noindent\rule{\linewidth}{0.4pt}
\vspace{-0.6em}

\begin{minipage}[c]{0.04\linewidth}
(b)
\end{minipage}%
\begin{minipage}[c]{0.96\linewidth}
\centering
\begin{tikzpicture}[
    >=Stealth, thick,
    embset/.style={draw, dashed, rounded corners=2pt, fill=orange!12,
        minimum width=1.4cm, minimum height=0.50cm, font=\small, align=center},
    distbox/.style={draw, rounded corners=3pt, fill=green!12,
        minimum width=1.6cm, minimum height=0.50cm, font=\small, align=center},
    result/.style={draw, rounded corners=3pt, fill=gray!12,
        minimum width=1.4cm, minimum height=0.50cm, font=\small, align=center},
]
\node[embset] (er) at (-2.5,  0.45) {$\mathcal{E}_{r,l}$};
\node[embset] (et) at (-2.5, -0.45) {$\mathcal{E}_{t,l}$};

\node[distbox] (d) at (0.3, 0) {$d(\cdot,\cdot)$};
\node[result] (dl) at (2.8, 0) {$d_l$};

\draw[->] (er.east) -- ([yshift= 4pt]d.west);
\draw[->] (et.east) -- ([yshift=-4pt]d.west);

\draw[->] (d.east) -- (dl.west);
\end{tikzpicture}
\end{minipage}

\caption{Overview of the applied method. (a) Layer-wise feature extraction from a
pretrained speech model. (b) Distance computation between reference and test
embedding distributions at layer \textit{l}.}
\label{fig:method}
\end{figure}

As the information contained in SFM representations depends on the model layer~\cite{pasad2021_layerwise_analysis_ssl}, in this work, we conduct a layer-wise analysis of SFM representations in order to determine the most suited representations for intelligibility prediction without any further training. We employ various distance measures to compare representations of reference and test signals.
Embedding distances have been successfully applied and compared in the audio domain \cite{kilgour19_interspeech_fad, biswas2026evaluatinggenerativeaudioinsights}. However, no works comparing multiple embedding distances for intelligibility prediction exist to our knowledge.
A layer-wise analysis of various SFMs was conducted in \cite{cooper2025layerwiseanalysis}, where correlations of the 2-Wasserstein distance with human quality and intelligibility ratings of multilingual synthesized speech were evaluated. 
These experiments are restricted to models' encoder layers and system-aggregated correlations, while we also include decoder layers where applicable, focus on signal-level correlations, i.e., higher granularity, and use enhanced and coded speech.
In \cite{kad}, a comparison of scaled \ac{MMD} and \ac{FAD} was done in the audio domain, demonstrating sample bias and computational cost as \ac{FAD} limitations.
An evaluation of \ac{FAD} and scaled \ac{MMD} with multiple audio embedding models and neural codec embeddings of coded audio signals at signal level was done by \cite{biswas2026evaluatinggenerativeaudioinsights}, demonstrating correlations with subjective MUSHRA \cite{ITU-BS1534-3-2015} results. However, only the last encoder layer of evaluated models was used.
Lastly, it has been shown that the evaluation of different model capacities has yielded different outcomes. In \cite{ravuri2024uncertaintyaspredictor}, adding dropout layers to a model resulted in higher correlations with subjective quality scores. Conversely, embeddings of more complex models showed better correlations with intelligibility in \cite{zezario_mosanetplus}. Motivated by this, we compare different model sizes.


In summary, the contributions of this work are:
\begin{itemize}
    \item[1)]We assess the applicability of representations of several SFMs for intrusive intelligibility prediction and find that Whisper representations are suited best.
    \item[2)] We conduct a layer-wise analysis of Whisper and find that the last encoder and decoder layers yield best results.
    \item[3)] We identify \ac{FAD} as the best-performing distance measure and show that distance measures on SFM representations are more robust across the evaluated datasets than baseline measures such as \ac{CER}. In contrast to non-differentiable baselines, distance measures are differentiable and may be used as a loss function.
    \item[4)] We compare different model sizes and show that correlations of intelligibility scores based on embedding distance with subjective intelligibility scores improve with model size.
\end{itemize}


\section{Methodology}



Fig.~\ref{fig:method} gives an overview of the applied method. We consider a pretrained SFM which maps speech signal frames $\speechFrame\in\setInp$ to a sequence of embedding vectors $\embedding_\idxLayer$. The embedding vectors $\embedding_\idxLayer$ are obtained as intermediate representations or the final output of the speech model
\begin{equation}
    \featmodel_\idxLayer \colon \setInp \to \mathbb{R}^{\featDim_\idxLayer},\quad \embedding_\idxLayer := \featmodel_\idxLayer(\speechFrame),
\end{equation}
where $\idxLayer$ indexes the SFM layer and $\featDim_\idxLayer$ is the the number of corresponding embedding dimensions. For using the pretrained SFM for speech intelligibility estimation, distances between embeddings of a reference set comprising clean speech $\setInp_{\mathrm{r}}$ (assuming perfect intelligibility) and corresponding test signals $\setInp_{\mathrm{t}}$ are computed. The layer-wise embeddings of reference and test signals are collected in the sets $\setEmb_{\mathrm{r},\idxLayer}= \{{\embedding_{\mathrm{r},i}^\idxLayer}\}_{i=1}^\numEmbR$ and $\setEmb_{\mathrm{t},\idxLayer}= \{{\embedding_{\mathrm{t},i}^\idxLayer}\}_{i=1}^\numEmbT$, where $\numEmbR$ and $\numEmbT$ are the numbers of samples in $\setEmb_{\mathrm{r},\idxLayer}$ and $\setEmb_{\mathrm{t},\idxLayer}$, respectively. 
The corresponding distributions are given by 
\begin{equation}
  \embedding_{\mathrm{r},i}^\idxLayer \sim \distR\ \ \text{for}\ \embedding_{\mathrm{r},i}^\idxLayer \in \setEmb_{\mathrm{r},\idxLayer}\quad \text{and}\quad \embedding_{\mathrm{t},i}^\idxLayer \sim \distT\ \ \text{for}\ \embedding_{\mathrm{t},i}^\idxLayer \in \setEmb_{\mathrm{t},\idxLayer}.  
\end{equation}
In the following, we will omit the layer index $\idxLayer$ whenever possible without adding confusion.

As the most simple distance, we employ the Euclidean distance (L2) between sample means
\begin{equation}
    \hat{\muRef}=\frac{1}{\numEmbR}\sum_{i=1}^{\numEmbR}\embedding_{\mathrm{r},i}\quad \text{and}\quad\hat{\muTest}=\frac{1}{\numEmbT}\sum_{i=1}^{\numEmbT}\embedding_{\mathrm{t},i}
\end{equation}
of the reference and test embeddings
\begin{equation}
    \mathrm{L2}(\hat{\muRef},\hat{\muTest}) = \Vert \hat{\muRef} - \hat{\muTest}\Vert_2.
\end{equation}
Similarly, we define the cosine distance (cosD) on sample mean vectors of embeddings
\begin{equation}
    \mathrm{cosD}(\hat{\muRef},\hat{\muTest}) = 1 - \frac{\hat{\muRef}^{\text{T}}\hat{\muTest}}{\Vert\hat{\muRef}\Vert_2 \ \Vert\hat{\muTest}\Vert_2}.
    \label{eq:cosd}
\end{equation}
%
%
%
To capture also complicated embedding structures and in particular also consider the spread of the data, we aim at measuring a distance in terms of embedding distributions rather than simple distances of embedding centroids. To this end, the $2$-Wasserstein distance between two Gaussian distributions \mbox{$\mathcal{N}(\muRef,\SigmaRef)\approx\distR$} and $\mathcal{N}(\muTest,\SigmaTest)\approx\distT$ that approximate the distribution of reference and test embeddings, respectively. We refer to this as the squared \ac{FAD}
\begin{equation}
  \FAD(\distR,\, \distT)
  = \FADformula,
  \label{eq:fad}
\end{equation}
where $\tr[\cdot]$ denotes the trace of a matrix and $\sqrt{\cdot}$ denotes the matrix square root here.
For practical application, $\muRef$, $\muTest$ and $\SigmaRef$, $\SigmaTest$ are estimated from $\setRef$ and $\setTest$ by sample mean vectors and sample covariance matrices, respectively.



A distribution-free, unbiased alternative is the \ac{MMD}, which has also been applied in the audio domain under the name of Kernel Audio Distance~\cite{kad}. The \ac{MMD} with respect to a positive definite kernel $\kernel$ between two distributions $\distR$ and $\distT$ is defined as
\begin{align}
  &\MMD^2(\distR, \distT) = \label{eq:mmd}\\ &\qquad\mathbb{E}_{\embedding_{\mathrm{r}}, \embedding_{\mathrm{r}'}}\!\left[\kernel(\embedding_{\mathrm{r}}, \embedding_{\mathrm{r}'})\right]
  + \mathbb{E}_{\embedding_{\mathrm{t}},\embedding_{\mathrm{t}}'}\!\left[\kernel(\embedding_{\mathrm{t}},\embedding_{\mathrm{t}}')\right]
  - 2\,\mathbb{E}_{\embedding_{\mathrm{r}},\embedding_{\mathrm{t}}}\!\left[\kernel(\embedding_{\mathrm{r}}, \embedding_{\mathrm{t}})\right],\notag
\end{align}
where $\embedding_{\mathrm{r}}, \embedding_{\mathrm{r}'}\sim\distR$ and $\embedding_{\mathrm{t}},\embedding_{\mathrm{t}}'\sim\distT$. 
An unbiased estimator of \eqref{eq:mmd} is 
\begin{align}
  &\widehat{\MMD}^2(\setRef, \setTest) = \frac{1}{\numEmbR(\numEmbR-1)} \sum_{i \neq j} \kernel(\embedding_{\mathrm{r},i}, \embedding_{\mathrm{r},j}) \label{eq:mmd_estimator}\\
  &\qquad + \frac{1}{\numEmbT(\numEmbT-1)} \sum_{i \neq j} \kernel(\embedding_{\mathrm{t},i}, \embedding_{\mathrm{t},j})
  - \frac{2}{\numEmbR\numEmbT}    \sum_{i=1}^{\numEmbR} \sum_{j=1}^{\numEmbT} \kernel(\embedding_{\mathrm{r},i}, \embedding_{\mathrm{t},j}),\notag
\end{align}

%
Following~\cite{kad}, we use a radial basis function kernel
\begin{equation}
  k(\embedding, \embedding')
    = \exp\!\left(-\frac{\|\embedding - \embedding'\|_2^2}{2\sigma^2}\right)
  \label{eq:rbf-kernel}
\end{equation}
and select the bandwidth $\sigma$ via the median-distance heuristic~\cite{JMLR:v13:gretton12a}.

By encoding a full speech utterance, the SFM $\featmodel$ will typically produce a sequence of embedding vectors at each layer. Hence, all discussed metrics may be computed either at system or signal level. 
We compute all metrics at signal level, where the embeddings of a single audio signal form an embedding set. For L2 and cosD, embeddings of a signal are averaged over the time dimension.
While FAD and MMD are traditionally used at system level by temporal averaging of each signal's embeddings and calculating the FAD and MMD over average embeddings of multiple signals, we perform signal-level computation following \cite{biswas2026evaluatinggenerativeaudioinsights}.


\section{Experimental Setup and Results}

\subsection{Datasets}
For evaluation, we use two datasets with human intelligibility responses. First, the NCLEIR (Noisy Coded Listening Effort and Intelligibility Responses) dataset~\cite{behringer2026assessingimpactnoisespeech}, consisting of clean, noisy, and noisy-then-enhanced English speech processed by six speech codecs. Since the codecs used in the NCLEIR dataset were trained to reconstruct speech rather than remove noise, we compare clean/noisy coded utterances to clean/noisy reference utterances, and noisy-then-enhanced coded utterances to noisy-then-enhanced ``reference" utterances. Second, the TMHINTQI VoiceMos2023 Track 3 Test set\footnote{\href{https://github.com/dhimasryan/TMHINT-QI-VoiceMOS2023}{github.com/dhimasryan/TMHINT-QI-VoiceMOS2023}}~\cite{voicemos_challenge_2023}, which contains noisy Mandarin speech, enhanced by five speech enhancement systems and without enhancement. Since TMHINTQI is focused on speech enhancement, we compare noisy and noisy-then-enhanced utterances to the clean reference.

For both datasets, the intelligibility score per utterance corresponds to the ratio of correctly transcribed words to total words. For correlation purposes, the scores for each utterance are averaged across listeners.

\subsection{Evaluated Features and Baseline Measures}
For evaluating the distance metrics, we use features extracted from three SFM models and a baseline of non-neural features.

\noindent\textbf{Whisper~\cite{radford2023robust}:} A weakly supervised multilingual multi-task model consisting of a Transformer encoder-decoder architecture, trained for speech recognition, spoken language identification, speech translation, and voice activity detection. We use the Tiny, Base, and Large-v3 models\footnote{\href{https://github.com/openai/whisper}{github.com/openai/whisper}}.

\noindent\textbf{wav2vec 2.0~\cite{NEURIPS2020_wav2vec2}:} A self-supervised model consisting of a Transformer encoder, trained for identifying quantized latent speech representations with a contrastive loss. We use the pretrained Base and the ASR-finetuned Base-960h models\footnote{\href{https://github.com/facebookresearch/fairseq/tree/main/examples/wav2vec}{github.com/facebookresearch/fairseq/tree/main/examples/wav2vec}} (abbreviated by W2V2-PT and W2V2-ASR, respectively in the following).

\noindent\textbf{WavLM~\cite{wavlm}:}
A self-supervised model consisting of a Transformer encoder, trained for discrete token prediction including a denoising task for improved robustness. We use the Base model\footnote{\href{https://github.com/microsoft/unilm/tree/master/wavlm}{github.com/microsoft/unilm/tree/master/wavlm}}.

\noindent\textbf{Mel-Frequency Cepstral Coefficients (MFCCs):} As non-neural baseline features, we compute 40 MFCCs at \SI{16}{kHz} sampling rate using \SI{25}{ms} window size, \SI{10}{ms} hop length, and a Mel filterbank with 80 bands.

For each SFM, we extract features from all Transformer layer outputs. For Whisper, this also includes decoder layers, which is in contrast to other works which only evaluate encoder layers~\cite{cooper2025layerwiseanalysis, zhou2025unveilingbestpracticesapplying}.


\noindent\textbf{Baseline Measures:} As baselines, we compute various intelligibility measures used in the literature. STOI~\cite{TaalSTOI} and  ESTOI~\cite{jensen2016_estoi} are employed as classical signal-processing metrics. Further, \ac{WER} and \ac{CER} are computed using Whisper. Since ground truth transcripts are not available for all test data, for computing the \ac{WER} and \ac{CER} of a test signal we treat the Whisper transcript of the corresponding reference signal as the ground truth. Lastly, as state-of-the-art measure with high computational complexity, we evaluate the neural prediction model MOSA-Net+~\cite{zezario_mosanetplus} (M+ in the following).
Given an input signal, M+ utilizes both acoustic and Whisper encoder features to predict speech quality and intelligibility. We use a provided checkpoint\footnote{\href{https://github.com/dhimasryan/MOSA-Net-Cross-Domain/tree/main/MOSA_Net\%2B}{github.com/dhimasryan/MOSA-Net-Cross-Domain/tree/main/MOSA\_Net\%2B}} trained on TMHINTQI with Whisper Large-v3 embeddings and evaluate only the intelligibility scores.
Since M+ is non-intrusive, we evaluate it using only the degraded signals also employed by the other measures to ensure a fair comparison.


\subsection{Model Comparison}



We first determine in which model's feature space embedding distances correlate most strongly with subjective intelligibility scores.
To ensure similar model complexities, we use the ``Base" models. Here, we compute the cosine distance (cosD, see \eqref{eq:cosd}) and correlate with subjective results using the \ac{PCC} and the \ac{SRCC}. Negative correlations are desired, which would indicate that smaller distances predict higher intelligibility.


\begin{table}[th]
\centering
\caption{\ac{PCC} and \ac{SRCC} for best-layer results across models for cosD. Best correlation per distance is printed in bold.}
\label{tab:model_comparison_signal_pearson}
\resizebox{\linewidth}{!}{%

\begin{tabular}{lrrrr}
\toprule
& \multicolumn{2}{c}{NCLEIR} & \multicolumn{2}{c}{TMHINTQI} \\
\cmidrule(lr){2-3} \cmidrule(lr){4-5}
Model Features & cosD PCC & cosD SRCC & cosD PCC & cosD SRCC \\
\midrule
MFCCs              & -0.218 & -0.133 & 0.214 & 0.096 \\
WavLM              & -0.073 & -0.119 & -0.312 & -0.326  \\
W2V2-PT            & -0.007 & 0.009  & 0.023 & -0.036 \\
W2V2-ASR           & -0.320 & -0.353 & -0.174 & -0.238 \\
Whisper Base       & \textbf{-0.560} & \textbf{-0.463} & \textbf{-0.583} & \textbf{-0.621} \\
\bottomrule
\end{tabular}}
\end{table}

The results are shown in Table~\ref{tab:model_comparison_signal_pearson}.  Whisper Base yields the strongest overall correlations for both datasets, with the best results obtained at the last decoder layer for NCLEIR and the last encoder layer for TMHINTQI. 
For the cosD of W2V2-ASR representations, moderate PCC and SRCC are found for NCLEIR, whereas only weak correlations are obtained for TMHINTQI. We attribute this to the English-only ASR fine-tuning data.
In contrast, W2V2-PT shows no correlation at all. The contrast between W2V2-PT and W2V2-ASR on NCLEIR indicates that the ASR objective improves the usefulness of the embeddings for predicting intelligibility.
MFCCs show a weak positive PCC for TMHINTQI, which is likely a random effect and not considered meaningful. Aside from Whisper and the non-correlated W2V2-PT, the best results for WavLM and W2V2-ASR are obtained at the last three encoder layers.

\subsection{Layer-Wise Evaluation}


\begin{figure*}[t]
    \centering
    \includegraphics[width=1\textwidth]{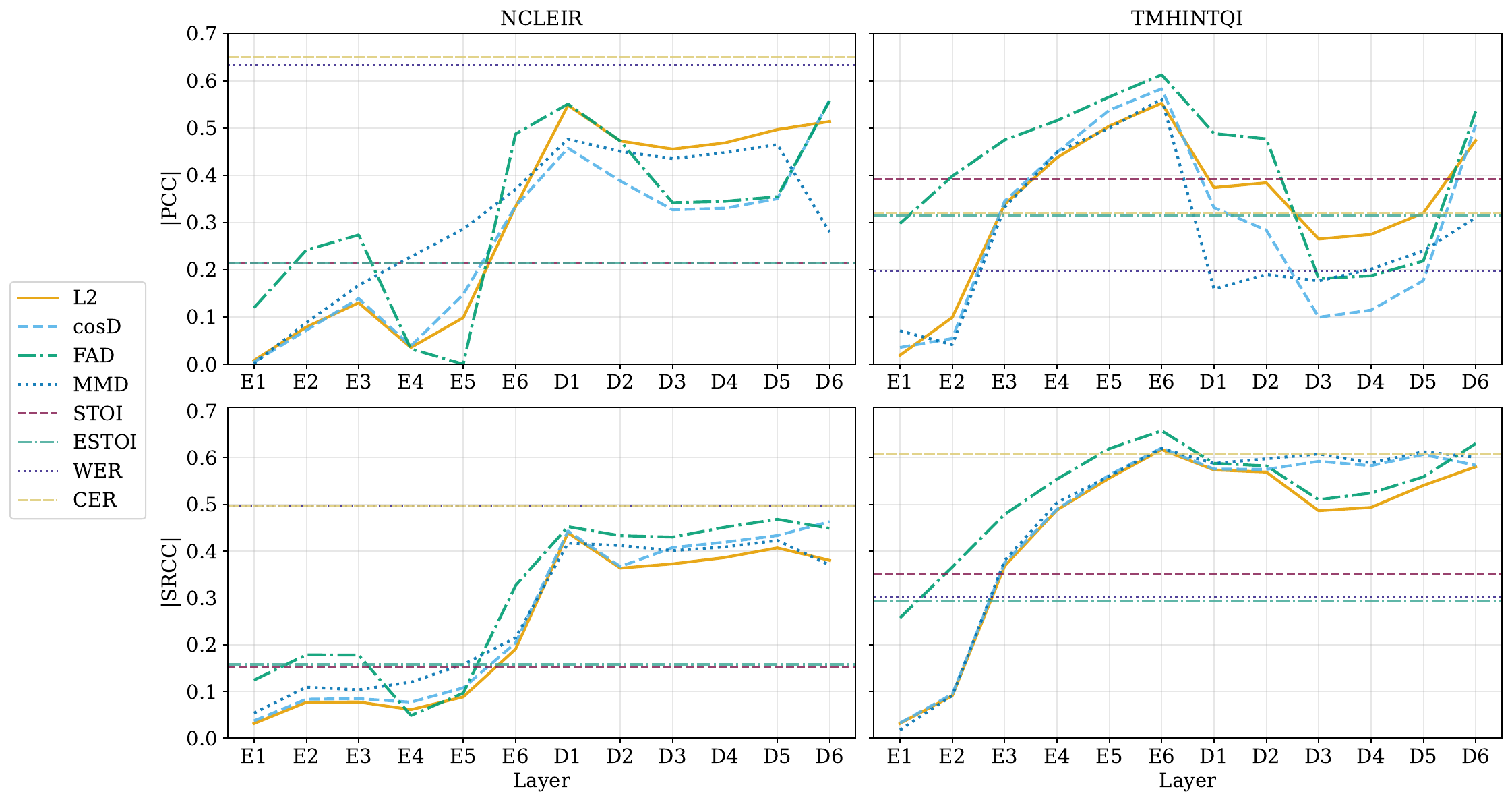}
    \caption{Absolute \ac{PCC} (top) and absolute \ac{SRCC} (bottom) with subjective intelligibility for layer-wise embedding distances shown for different distances as well as baseline measures (E = encoder layer, D = decoder layer).}
    \label{fig:whisper_base_distance_corrs}
\end{figure*}

As Whisper features performed best in the model comparison, we evaluate this model more closely. We conduct a layer-wise evaluation of Whisper Base and compare the distance metrics in terms of \ac{PCC} and \ac{SRCC} with subjective results of NCLEIR and TMHINTQI. \ac{WER} and \ac{CER} are also computed using the Whisper Base model.

Figure~\ref{fig:whisper_base_distance_corrs} shows the layer-wise absolute\footnote{The absolute values are used for visualization purposes. All signed correlation values $< -0.1$ or $> 0.1$ align with the expected correlation direction, i.e., positive for STOI/ESTOI and negative for all other measures.} \ac{PCC} and \ac{SRCC} for Whisper Base on the NCLEIR and TMHINTQI datasets. 
We first discuss \ac{PCC}, then \ac{SRCC}.
Among the distances evaluated on NCLEIR, the strongest \ac{PCC} values are given by cosD and FAD at the last decoder layer. Slightly lower \ac{PCC} is found for L2 and MMD at the first decoder layer. Surprisingly, despite a limited number of embedding frames per signal, especially in the decoder, FAD does not show numerical issues in practice and outperforms the unbiased MMD. STOI and ESTOI show only weak \ac{PCC}, while the \ac{PCC} of \ac{CER} and \ac{WER} is greater than that all embedding distances. An inspection of the \ac{WER} and \ac{CER} scores revealed that they are mostly below 0.1, with  remaining scores gradually declining towards 1.0, demonstrating that Whisper is capable of accurately transcribing the majority of NCLEIR test signals.
For TMHINTQI, the strongest correlations for all embedding distances are found at the last encoder layer, with FAD showing the strongest overall correlation. Notably, all embedding distances yield higher absolute correlations than the \ac{WER} and \ac{CER} for this dataset. Here, \ac{WER} scores are  predominantly 1.0 and \ac{CER} is spread between 0 and 1.2. While \ac{WER} is generally not recommended for Mandarin data due to the lack of word boundaries \cite{k-etal-2025-advocating-CER}, the worse \ac{CER} results compared to NCLEIR show that the final ASR output is less reliable for Mandarin. This highlights the increased robustness of embeddings compared to discrete outputs from the same SFM. 

The results for \ac{SRCC} largely follow the same trends as for \ac{PCC}. A notable difference is the \ac{CER} for TMHINTQI, which is comparable to the embedding distances, but still outperformed by the FAD at the last encoder layer. This indicates that the relationship between \ac{CER} and subjective intelligibility is monotonic but non-linear.
Moreover, the embedding distances show substantial drops in \ac{PCC} at the intermediate decoder layers, while for \ac{SRCC}, only a slight decrease on the TMHINTQI data is found for FAD and L2.

Overall, FAD is the most robust distance, with the final encoder and decoder layers yielding the best results. While the best-ranked layer varies depending on the test set, FAD using the last decoder layer yields the overall best correlations across datasets.

\subsection{Impact of Whisper Model Size}
As the last experiment, we evaluate the impact of Whisper's model size on \ac{PCC}. We compare Whisper Tiny, Base, and Large-v3 in terms of \ac{WER}, \ac{CER}, M+, and the embedding distances. The Tiny model has 384-dimensional embeddings and four encoder and decoder layers, Base has 512-dimensional embeddings and six respective layers, while Large-v3 has 1280-dimensional embeddings and 32 respective layers. Tiny and Base were trained on 680,000 hours of data, while Large-v3 was trained using 5 million hours of data.

\begin{table}
\caption{\ac{PCC} with subjective intelligibility scores for different metrics and Whisper model sizes (NC = NCLEIR, TM = TMHINTQI, best results are printed in bold).}
\label{tab:whisper_comparison}
\resizebox{\linewidth}{!}{%
\begin{tabular}{l|c|r|r|r|r|r|r|r}
\toprule
Model & Data & cosD $\downarrow$ & FAD $\downarrow$ & MMD $\downarrow$ & L2 $\downarrow$ & WER $\downarrow$ & CER $\downarrow$ & M+ $\uparrow$ \\
\midrule
Tiny & NC & -0.575 & -0.572 & -0.420 & -0.546 & -0.621 & -0.626 & - \\
Base & NC & -0.560 & -0.558 & -0.476 & -0.548 & -0.634 & -0.650 & - \\
Large-v3 & NC & \textbf{-0.634} & \textbf{-0.676} & \textbf{-0.585} & \textbf{-0.590} & \textbf{-0.720} & \textbf{-0.719} & \textbf{0.684} \\
\midrule
Tiny  & TM & -0.454 & -0.518 & -0.428 & -0.439 & -0.151 & -0.399 & - \\
Base  & TM & -0.583 & -0.613 & -0.562 & -0.553 & -0.198 & -0.320 & - \\
Large-v3 & TM & \textbf{-0.746} & \textbf{-0.725} & \textbf{-0.731} & \textbf{-0.688} & \textbf{-0.320} & \textbf{-0.657} & \textbf{0.770} \\
\bottomrule
\end{tabular}%
}
\end{table}

Table~\ref{tab:whisper_comparison} shows the results of the layers with highest correlation of the embedding distances for Whisper Tiny, Base and Large-v3 on NCLEIR and TMHINTQI. Similarly to the Base model, highest correlation is obtained for Tiny and Large-v3 at the first and last decoder layers for NCLEIR and the last encoder layer for TMHINTQI. Whisper Large-v3 shows the strongest correlations across all metrics. Base is mostly more correlated than Tiny across both datasets. M+ has the overall strongest correlation on the TMHINTQI test set, which is expected since it was trained on the TMHINTQI train set. For NCLEIR, M+ shows a correlation with the subjective intelligibility scores that is similar to FAD. Notably, M+ is outperformed by WER/CER, indicating that WER/CER might be preferable if the test data is expected to be more similar to the Whisper training data, i.e., predominantly English. The results indicate that using a larger model can yield more informative representations for intelligibility. However, this comes at the cost of higher model complexity.



\section{Conclusion}
In this work, we evaluated the usefulness of embedding distances using pretrained SFM representations for intelligibility prediction without any further training.
The results show that Whisper is the most suitable SFM for this task, FAD is the distance with best overall performance, and the best layers for feature extraction are the last encoder and decoder layers. Embedding distances calculated based on larger Whisper models are shown to yield stronger correlations.
Importantly, the correlations of the evaluated distances with subjective intelligibility scores are stronger than for signal-processing based metrics and more robust across datasets than \ac{WER} and \ac{CER}. Further, the embedding distances are differentiable enabling their use as intelligibility-based loss functions in neural network trainings, which is not possible for \ac{WER} and \ac{CER}. 
Surprisingly, despite expected lower sample efficiency, FAD does not show numerical issues in practice. 
While a more in-depth analysis on potential reasons was out of scope for this work, future work should investigate FAD more closely in terms of sample bias, the accuracy of covariance estimation, and their impact on intelligibility prediction.

\bibliographystyle{IEEEbib}
\bibliography{mybib}

\end{document}